\documentclass[10pt]{article}
\usepackage[english]{babel}									
\usepackage[utf8]{inputenc}				
\usepackage[T1]{fontenc}									
\usepackage{amsmath,amsfonts,amssymb,amsthm,cancel,siunitx,
calculator,calc,mathtools,empheq,latexsym}
\usepackage{subcaption}
\usepackage{longtable}
\usepackage{epsfig,tikz,float}		             
\usepackage{booktabs,multicol,multirow,tabularx,array}          
\usepackage{cite} 
\title{\large\bf%
\uppercase{Application of INFERNO to a top pair cross section measurement \\ with CMS Open Data}
}
\author{%
Lukas Layer$^{1,2}$, \ Tommaso Dorigo$^{1,3}$, and \ Giles Strong$^{1}$
}

\begin{document}

\date{}

\maketitle

\vspace{-0.5cm}

\begin{center}
{\footnotesize 
$^1$ INFN, Sezione di Padova \\
$^2$ Università di Napoli ``Federico II"\\
$^3$ Universal Scientific and Research Network
}
\end{center}

\bigskip
\noindent
{\small{\bf ABSTRACT.}
In recent years novel inference techniques have been developed based on the construction of non-linear summary statistics with neural networks by minimising inference-motivated losses.  One such technique is \textsc{inferno} (P. de Castro and T. Dorigo, Comp.\ Phys.\ Comm.\ 244 (2019) 170) which was shown on toy problems to outperform classical summary statistics for the problem of confidence interval estimation in the presence of nuisance parameters. In order to test and benchmark the algorithm in a real world application, a full, systematics-dominated analysis produced by the CMS experiment, "Measurement of the $\mathrm{t}\bar{\mathrm{t}}$ production cross section in the $\tau$+jets channel in pp collisions at $\sqrt{s}=7\ \mbox{TeV}$" (CMS Collaboration, The European Physical Journal C, 2013) is reproduced with CMS Open Data. The application of the \textsc{inferno}-powered neural network architecture to this analysis demonstrates the potential to reduce the impact of systematic uncertainties in real LHC analyses.
This work also exemplifies the extent to which LHC analyses can be reproduced with open data.
}

\baselineskip=\normalbaselineskip

\section{Introduction}\label{sec:0}

Over the course of the past decade, machine-learning-powered classification and regression models have become very popular in High Energy Physics (HEP) to construct powerful summary statistics that are used for inference. Despite their success, these methods rely on minimisation of a standard cross-entropy loss and on standard measures of performance for the learning task, neither of which are aligned with the inference goal when observations depend on nuisance parameters. The presence of nuisance parameters then causes a reduction of the statistical power of the summary statistics during inference. 

In recent years, a novel approach, called \textsc{inferno}~\cite{DECASTRO2019170}, an acronym that stands for Inference-Aware Neural Optimization, has been developed to construct machine-learning-based summary statistics that are optimal for the specific analysis goal. The authors proposed a method to construct non-linear summary statistics by minimising inference-motivated losses via stochastic gradient descent. 
The proposed algorithm can be used to directly minimise an approximation of the expected uncertainty on the parameter of interest, fully accounting for the effect of considered nuisance parameters.
In the mentioned work the algorithm was tested with a synthetic example, inspired by the typical conditions of a cross section measurement. It was shown that the confidence intervals obtained using \textsc{inferno}-based summary statistics are narrower than those using binary classification, and they tend to be closer to those expected when using the true model likelihood for inference. The improvement over binary classification was also shown to increase when more nuisance parameters are considered.
In spite of those promising results, the use of a synthetic dataset for performance tests left unclear the issue of how the algorithm performs when confronted with the complications of a full-fledged LHC analysis. In this work we fill that gap, by describing the development of a framework where real LHC data can be used to train with the \textsc{inferno} algorithm. In order to benchmark the framework, a full, systematics-dominated analysis of the CMS experiment, "Measurement of the $\mathrm{t}\bar{\mathrm{t}}$ production cross section in the $\tau$+jets channel in pp collisions at $\sqrt{s}=7\ \mbox{TeV}$"~\cite{ttjets} is reproduced with CMS Open Data. 
The code for this study has been released to the  public~\cite{lukas_layer_2022_6080791}.

\section{Inference Aware Neural Optimization}\label{sec:1}

The \textsc{inferno} algorithm~\cite{DECASTRO2019170} aims at directly minimising the expected variance of the parameter of interest (POI) obtained via a non-parametric simulation-based synthetic likelihood.
\begin{figure}[h]
  \centerline{\includegraphics[width=1.\textwidth]{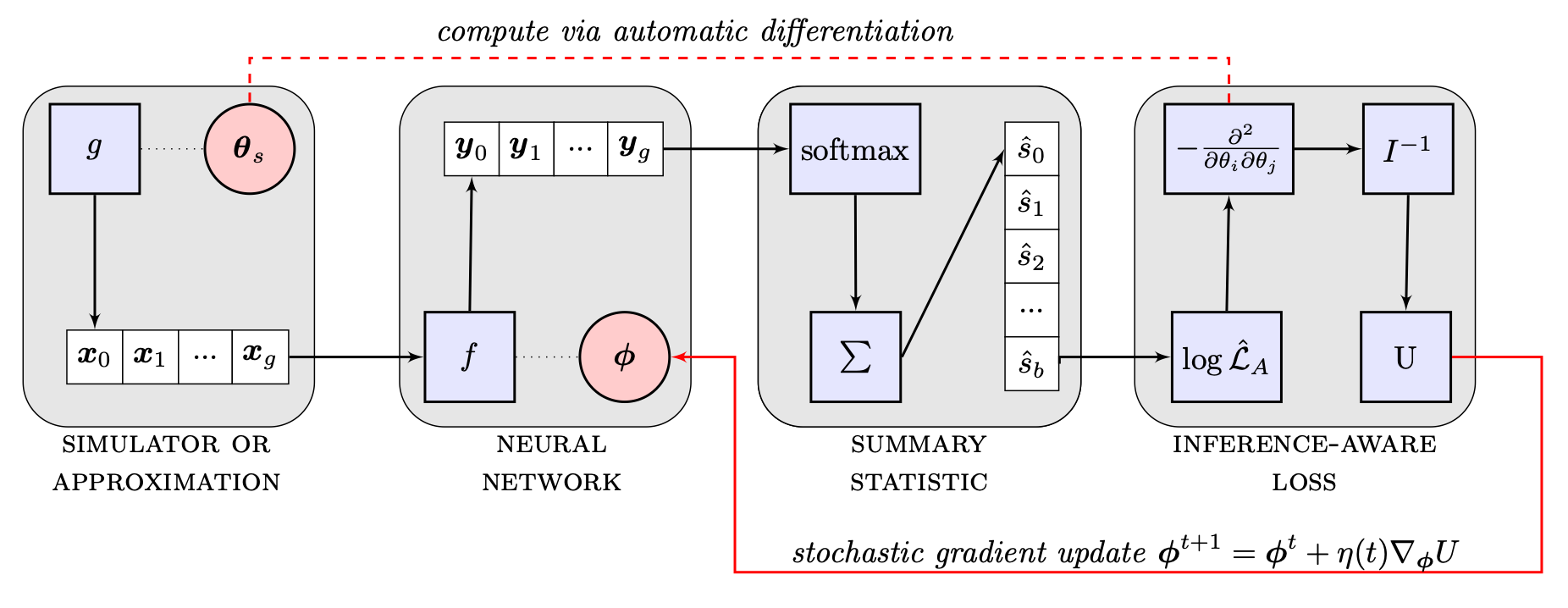}}
  \caption{Sketch of the \textsc{inferno} algorithm. Batches from a simulator are passed through a neural network and a differentiable summary statistic is constructed that allows one to calculate the variance of the POI. The parameters of the network are then updated by stochastic gradient descent (SGD). The figure is reproduced with permission from~\cite{DECASTRO2019170}.}
  \label{inferno_sketch}
\end{figure}
The parameters of a neural network are optimised by stochastic gradient descent via automatic differentiation, where the loss function accounts for the details of the statistical model and in particular for the effect of nuisance parameters.
The original algorithm has been implemented in \textsc{\hbox{TensorFlow 1}}~\cite{tensorflow2015-whitepaper}.
A sketch of the \textsc{inferno} algorithm is shown in Fig.~\ref{inferno_sketch}.
An inference-aware summary statistic is learnt by optimising the parameters $\boldsymbol{\phi}$ of a neural network $f$ in order to reduce the dimensionality $d$ of each input observation $\boldsymbol{x}$:
\begin{equation}
f(\boldsymbol{x} ; \boldsymbol{\phi}):  \mathbb{R}^{d} \rightarrow \mathbb{R}^{b} ~.
\end{equation}
The network is trained with batches of simulated samples $G_{s}=\left\{\boldsymbol{x}_{0}, \ldots, \boldsymbol{x}_{g}\right\}$ obtained from a simulator $g$ with parameters $\boldsymbol{\theta}_{s}$. Here $G_s$ denotes one batch of samples. The number of nodes in the last layer of the network determines the dimension $b$ of the summary statistic. Since histograms are not differentiable, the original algorithm uses a softmax function as a differentiable approximation for the neural network output $y$:
\begin{equation}
\hat{s}_{i}(G_{s} ; \boldsymbol{\phi})=\sum_{\boldsymbol{x}} \frac{e^{f_{i}(\boldsymbol{x} ; \phi) / \tau}}{\sum_{j=0}^{b} e^{f_{j}(\boldsymbol{x} ; \boldsymbol{\phi}) / \tau}}
\end{equation}
where the temperature hyperparameter $\tau$ regulates the softness of the operator. For small temperatures $\tau \rightarrow 0^{+}$, the probability of the largest component will tend to $1$ while others to $0$. With this approximation it is possible to construct a summary statistic for each batch by computing the Asimov Poisson-count likelihood\footnote{ The concept of Asimov data is also referred to as a “saturated model” in Statistics.} $\hat{\mathcal{L}}_{A}$:
\begin{equation}
\hat{\mathcal{L}}_{A}(\boldsymbol{\theta} ; \boldsymbol{\phi})=\prod_{i=0}^{b} \operatorname{Pois}\left(\hat{s}_{i}\left(G_{s} ; \boldsymbol{\phi}\right) | \hat{s}_{i}\left(G_{s} ; \boldsymbol{\phi}\right)\right)~.
\end{equation}
The maximum-likelihood estimate (MLE) for the Asimov likelihood is the parameter vector $\boldsymbol{\theta}_{s}$ used to generate the simulated dataset $G_{s}$, {\em i.e.}\ $\operatorname{argmax}_{\boldsymbol{\theta}}\left(\hat{\mathcal{L}}_{A}(\boldsymbol{\theta} ; \boldsymbol{\phi})\right)=\boldsymbol{\theta}_{s}$.  The effect of the parameters of interest and the main nuisance parameters can be included by changing the mixture coefficients of mixture models, translations of a subset of features, or conditional density ratio re-weighting. An example for this will be discussed in the next section, where the application of the algorithm 
is described.
From the Asimov likelihood the Fisher information matrix is then calculated via automatic differentiation according to:
\begin{equation}
\boldsymbol{I}(\boldsymbol{\theta})_{i j}=\frac{\partial^{2}}{\partial \theta_{i} \partial \theta_{j}}\left(-\log \hat{\mathcal{L}}_{A}(\boldsymbol{\theta} ; \boldsymbol{\phi})\right)~.
\end{equation}
The covariance matrix can be estimated from the inverse of the Fisher information matrix if $\hat{\boldsymbol{\theta}}$ is an unbiased estimator of the values of $\boldsymbol{\theta}$:
\begin{equation}
\operatorname{cov}_{\boldsymbol{\theta}}(\hat{\boldsymbol{\theta}}) \geq I(\boldsymbol{\theta})^{-1} ~.
\end{equation}
It is also possible to include auxiliary measurements that constrain the nuisance parameters, characterized by likelihoods $\left\{\mathcal{L}_{C}^{0}(\boldsymbol{\theta}), \ldots, \mathcal{L}_{C}^{c}(\boldsymbol{\theta})\right\}$,
by considering the augmented likelihood $\hat{\mathcal{L}}_{A}^{\prime}$:
\begin{equation}
\hat{\mathcal{L}}_{A}^{\prime}(\boldsymbol{\theta} ; \boldsymbol{\phi})=\hat{\mathcal{L}}_{A}(\boldsymbol{\theta} ; \boldsymbol{\phi}) \prod_{i=0}^{c} \mathcal{L}_{C}^{i}(\boldsymbol{\theta}) ~.
\end{equation}
The loss function used to optimise the parameters of the neural network $\boldsymbol{\phi}$ can be any function of the covariance matrix at $\boldsymbol{\theta}_{s}$, depending on the concrete inference problem being studied. 
The diagonal elements $I_{i i}^{-1}\left(\boldsymbol{\theta}_{s}\right)$ correspond to the expected variance for the parameter $\theta_{i}$. Thus, if the aim is optimal inference on one of the parameters $\omega_{0}=\theta_{k}$, a possible loss function is:
\begin{equation}
U=I_{k k}^{-1}\left(\boldsymbol{\theta}_{s}\right)
\end{equation}
which corresponds to the approximated expected width of the confidence interval for $\omega_{0}$.

\section{Replication of the $\mathrm{t}\bar{\mathrm{t}}$ production cross section measurement in the $\tau$+jets channel 
}\label{sec:2}

In order to convince large collaborations consisting of several thousand members of the usefulness and correctness of novel analysis methods, it is beneficial to test and benchmark the algorithms on datasets that are as realistic as possible. This motivates the development of a dataset that reproduces a full CMS analysis and is accessible by the public in order to facilitate comparisons between novel approaches to inference.
Therefore this work reproduces a published CMS analysis, 
based on real \mbox{Run 1} legacy data of the CMS experiment, available in the CERN Open Data portal~\cite{opendata2021}. The reproduced analysis was first published by the CMS experiment in 2013. It measures the $\mathrm{t}\bar{\mathrm{t}}$ production cross section in the $\tau$+jets channel in pp collisions at $\sqrt{s}=7\ \mbox{TeV}$~\cite{ttjets}. 
\begin{figure}[h]
\begin{subfigure}[c]{0.5\linewidth}
\centering
\includegraphics[width=1\linewidth]{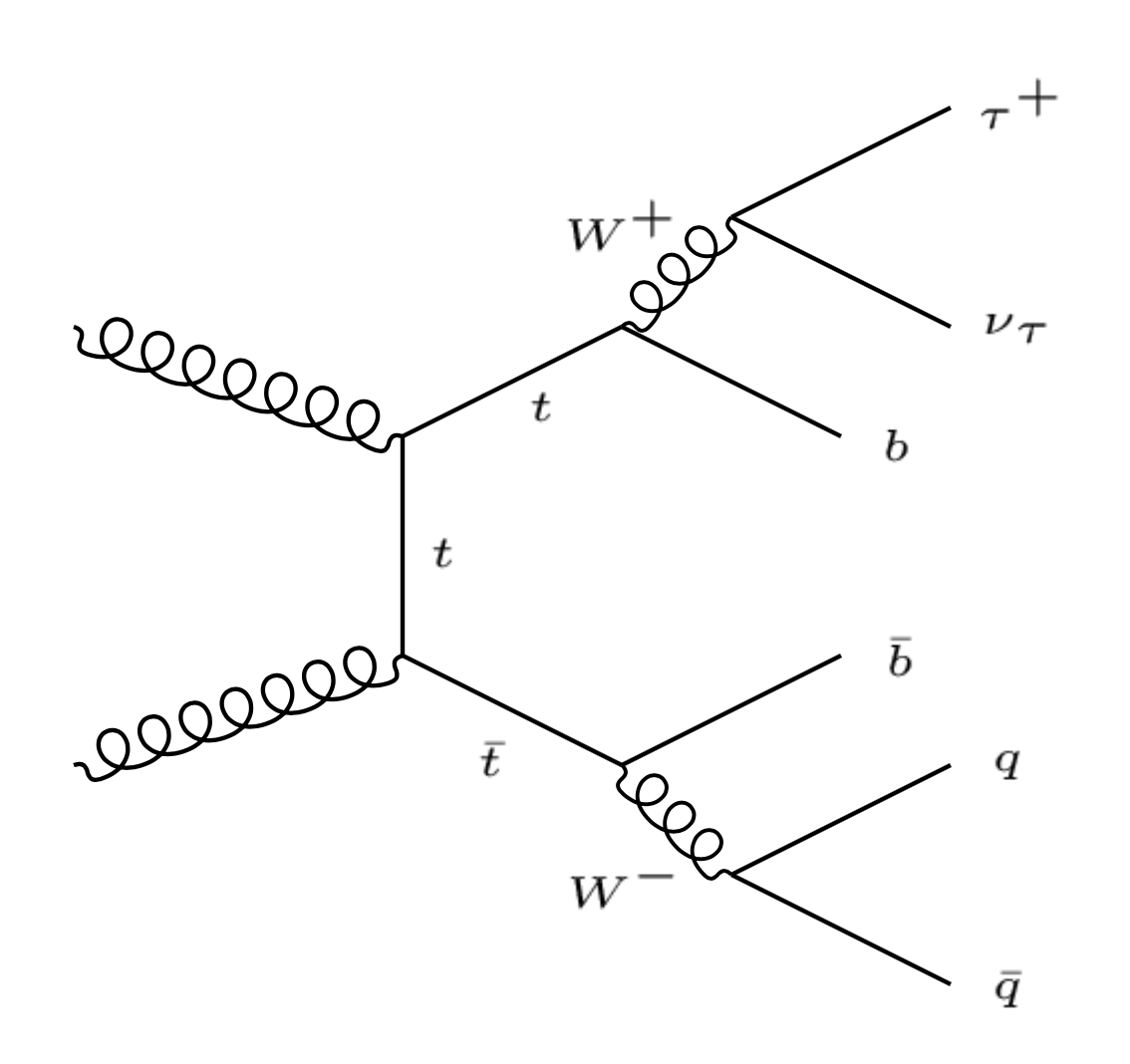}
\end{subfigure}
\hfill
\begin{subfigure}[c]{0.5\linewidth}
\centering
\includegraphics[width=1\linewidth]{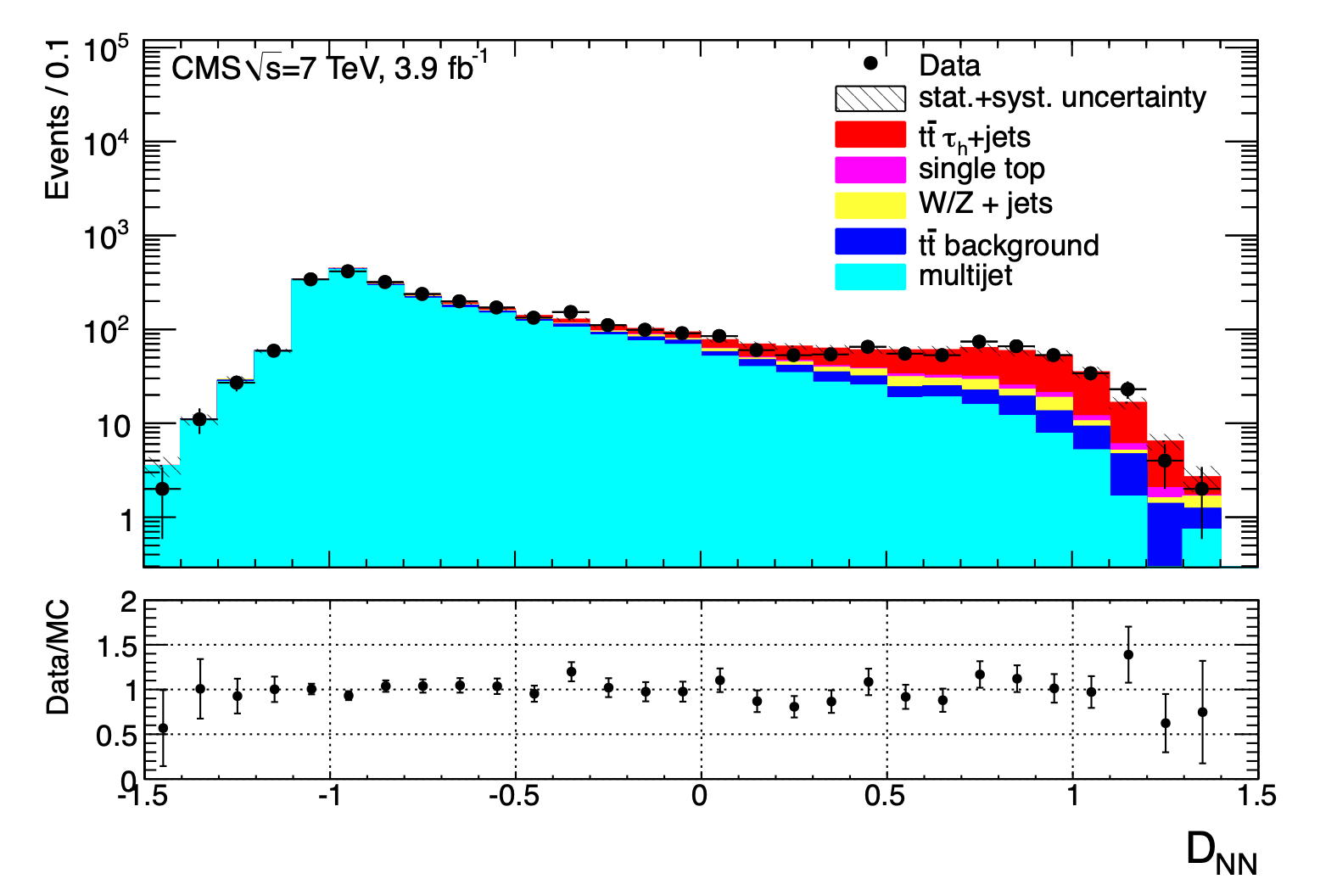}
\end{subfigure}
\caption{Left panel: Feynman diagram for the decay of a top pair into $\tau$+jets. Right panel: data-simulation agreement for the neural network classifier used in the original analysis~\cite{ttjets}.}
\label{fey}       
\end{figure}
In the context of testing and benchmarking the \textsc{inferno} algorithm, this analysis is of interest because its results are dominated by systematic uncertainties, and because it uses a neural network classifier to construct a summary statistic as input for the inference. Thus it constitutes a use case in which the \textsc{inferno} technique can possibly improve the precision of the measurement. \\
From a physics point of view, the measurement of the $\mathrm{t}\bar{\mathrm{t}}$ production cross section constitutes an important test of the Standard 
Model (SM), since the top quark plays a crucial role in many extensions of the SM due to its high mass. 
A direct measurement of the $\mathrm{t}\bar{\mathrm{t}}$ cross section in the $\tau$+jets final state offers the opportunity to investigate possible mass- or flavour-dependent couplings of the top quark. Moreover, a charged Higgs boson could give rise to an enhanced top pair cross section in proton-proton collisions. As depicted in the Feynman-diagram in the left panel of Fig.~\ref{fey}, the final state of this decay contains four hadronic jets, of which two are originated by bottom quarks, plus a hadronically decaying tau and a tau neutrino. 
The $\mathrm{t}\bar{\mathrm{t}}$ production cross section at $\sqrt{s}=7\ \mbox{TeV}$ measured in the original CMS analysis is:
$\sigma_{t t}=152 \pm 12 ~\mathrm{(stat.)} \pm 32 ~\mathrm{(syst.)} \pm 3 ~\mathrm{(lum.)~pb}$, which is consistent with the SM prediction. 
The data-simulation agreement for the neural network classifier is shown in the right panel of Fig.~\ref{fey}. 

\subsection{Datasets and Trigger}

The choice of the data and simulation samples follows the choices done in the original CMS analysis. A multijet trigger has been used to record $\mathrm{p p} \rightarrow \mathrm{t}\bar{\mathrm{t}} \rightarrow \tau_{\mathrm{h}}+\mathrm{j e t s}$ events. The trigger requires the presence of four calorimeter jets,  one of them identified as a tau lepton. Since the event rate increased during data taking with the rising instantaneous luminosity, two versions of the trigger have been developed:  \textsc{QuadJet40\_IsoPFTau40} and \textsc{QuadJet45\_IsoPFTau45}, where in the latter the $p_\mathrm{T}$ thresholds for the jets and the tau lepton were raised.
The used multijet triggers are part of the \textit{MultiJet} primary dataset collected by CMS in \mbox{Run 1}. 
The 2011 RunA and RunB  \textit{MultiJet} datasets have been released to the public and are available in the CMS Open Data database. Table \ref{sel_data} summarises the chosen trigger for each data taking period and the corresponding integrated luminosity.
\begin{table}
\begin{center}
\def\arraystretch{1.5}
\begin{tabular}{llll}
\hline Dataset & run range & trigger & $\mathcal{L}\left(\mathrm{pb}^{-1}\right)$ \\
\hline \hline Run2011A~\cite{cmsopenmultiruna} & $160431 - 165969$ & QuadJet40\_IsoPFTau40 & 357.5  \\
\hline Run2011A~\cite{cmsopenmultiruna} & $165970 - 166782$ & QuadJet45\_IsoPFTau45 & 363.5  \\
\hline Run2011A~\cite{cmsopenmultiruna} & $166783 - 171049$ & QuadJet40\_IsoPFTau40 & 514.7  \\
\hline   
\hspace{-.35cm}
\begin{tabular}{l}
Run2011A~\cite{cmsopenmultiruna} \\ Run2011B~\cite{cmsopenmultirunb}
\end{tabular}
 & $171050 - 178420$ & QuadJet45\_IsoPFTau45 & 2930.2 \\
\hline
\hline Total Luminosity &  &  & 4165.9  \\
\end{tabular}
\caption{Datasets with the chosen trigger, corresponding run numbers and luminosity. The version of the datasets is 12Oct2013-v1.}
\label{sel_data}
\end{center}
\end{table}
The total integrated luminosity of the dataset analysed with CMS Open Data is $4.16~\mathrm{fb}^{-1}$.
The trigger efficiencies have been recalculated with the CMS Open Data datasets, by separately measuring the efficiency of a single jet and a single hadronically decaying tau lepton to pass the trigger requirements. Following the original analysis, the trigger efficiency is modeled in simulation by multiplying the efficiencies obtained for the three most energetic central jets and the trigger efficiency obtained for the tau candidate. \\
\\
The legacy CMS Open Data Monte-Carlo (MC) simulation, called "Summer11 simulation", is used to estimate the signal efficiency as well as the contribution from electroweak background processes~\cite{cmsopenttjets, cmsopenwjets, cmsopendyjets, cmsopensttchan, cmsopensttbartchan, cmsopenstttwchan, cmsopensttbartwchan, cmsopenstschan, cmsopenstbarschan}. The $\mathrm{t}\bar{\mathrm{t}}$ signal and background events and the W/Z + jets events are simulated with the \textsc{\hbox{madgraph}} generator~\cite{madgraph2011} using the parton distribution function set CTEQ66~\cite{cteq2010}. The parton showering, fragmentation, hadronization
and decays of short lived particles, except tau leptons, is simulated with \textsc{\hbox{pythia}}~\cite{Sjostrand:2014zea}. Tau leptons are decayed using \textsc{\hbox{tauola}}~\cite{tauola}. Single-top events are simulated with \textsc{\hbox{powheg}}~\cite{powheg} interfaced to \textsc{\hbox{pythia}} and \textsc{\hbox{tauola}}. The used top-quark mass value is 172.5 GeV and the Next-to-Next-Leading-Log (NNLL) $\mathrm{t}\bar{\mathrm{t}}$ cross section is assumed to be 164 $\pm$ 10 pb~\cite{ttxsec}. 

\subsection{Event selection}

The event selection follows closely the original analysis~\cite{ttjets} and requires the presence of at least four particle-flow jets, and the presence of one particle-flow tau-lepton candidate.
The jets are reconstructed with the anti-$\mathrm{k_T}$ clustering algorithm~\cite{Cacciari:2008gp} with a distance parameter $R=0.5$. 
Selected events are required to have at least four particle-flow jets with pseudorapidity $|\eta| < 2.4$. Jets overlapping with leptons within $\Delta R(\mathrm{jet}, \mathrm{lepton}) > 0.4$ are excluded. To be consistent with the trigger, three jets are required to have $p_\mathrm{T} >45$ GeV and the fourth jet is required to have $p_\mathrm{T} > 20$ GeV. The jet candidates are required to be matched to jet objects used in the trigger within $\Delta R < 0.4$. 
The selected events are required to contain at least one b-tagged jet, identified with the recommended Combined Secondary Vertex algorithm (CSV) at its medium working point~\cite{btag2013}. 
The transverse missing energy (MET) is obtained with the particle-flow algorithm. A selection on the transverse missing energy, \mbox{$\mathrm{MET} > 20$ GeV}, is applied to reject QCD background.  \\
\\
The hadronically decaying tau-lepton candidate is reconstructed with the hadron-plus-strip (HPS) algorithm~\cite{hps2016}. 
The selected tau candidates have to fulfil an isolation criterion: the sum of the transverse energies of the additional charged hadrons and photons reconstructed in an isolation cone of $\Delta R=\sqrt{(\Delta \eta)^{2}+(\Delta \phi)^{2}}=0.5$ around the tau candidate is required to be less than $1~\mathrm{GeV}$. Furthermore, tau candidates are required to pass discriminators against muons and electrons.
The leading track of the tau candidate is vetoed if it is identified as a muon in order to suppress the muon contamination. In addition, the charged tau candidate may not be identified as a minimum ionising particle, therefore the ratio of the sum of the energy deposits in the ECAL and HCAL calorimeters associated to the tau candidate over the leading track momentum is required to be larger than $0.2$.
To be consistent with the trigger conditions, the transverse momentum of the tau candidate is required to fulfil $p_\mathrm{T} > 45$ GeV and the tau candidate is required to be matched within $\Delta R < 0.4$ to the tau object used in the trigger.\\
\\
In order to suppress the misidentification of electrons and muons as tau candidates, a veto on the presence of loosely isolated electrons and muons is applied. The isolation requirement is defined as $I / p_{\mathrm{T}}<0.15$, where $I$ is the sum of the transverse energy deposits in the ECAL and HCAL calorimeters and $p_{\mathrm{T}}$ is the scalar value of the track momenta within a cone of $\Delta R=0.3$.\\
\\
The dominating background in this analysis is constituted by high-jet-multiplicity (multijet) events from Quantum Chromo-Dynamical (QCD) processes, where one of the jets is misidentified as a hadronically-decaying tau lepton. 
The multijet background is estimated with a data-driven approach by inverting the b-tagging requirement, {\em i.e.}\ vetoing the presence of a b-tagged jet selected with the CSV algorithm. 
The smaller contributions from the electroweak processes are estimated from simulated events and are normalised to the theoretical cross section and the total integrated luminosity. 

\subsection{The Multivariate Classifier}

A neural network classifier is trained to discriminate simulated $\mathrm{t}\bar{\mathrm{t}}$ signal events from simulated QCD background events, yielding the final discriminant variable used as input to the statistical analysis. Feature engineering based on kinematic variables of the selected jets, tau-lepton and missing transverse energy is applied in order to construct high-level features that allow to better discriminate between signal and background. The following variables have been calculated and form the input for the multivariate classifier:
\begin{center}
\def\arraystretch{1.5}
\begin{longtable}{p{2.5cm}p{10cm}}
\hline Variable & Description \\
\hline 
\hline
$\mathrm{H}_T$  & scalar sum of the transverse momenta of all the selected jets and hadronic tau-lepton candidate \\
\hline
aplanarity & $A=\frac{3}{2} \lambda_{1}$ with $\lambda_{1}$ being the smallest eigenvalue of the momentum tensor $M^{\alpha \beta}=\sum_{i} p_{i}^{\alpha} p_{i}^{\beta} / \sum_{i}\left|\vec{p}_{i}\right|^{2}$, where $i$ runs over the number of jets and the tau candidate  and $\alpha, \beta=1,2,3$ specify the three spatial components of the momentum.\\
\hline
sphericity & $A=\frac{3}{2}\left(\lambda_{1}+\lambda_{2}\right)$ with $\lambda_{1}, \lambda_{2}$ being the smallest eigenvalue of the momentum tensor $M^{\alpha \beta}=\sum_{i} p_{i}^{\alpha} p_{i}^{\beta} / \sum_{i}\left|\vec{p}_{i}\right|^{2}$, where $i$ runs over the number of jets and the tau candidate. \\
\hline
$q \times \left|\eta\left(\tau_{\mathrm{h}}\right)\right|$ & charge of the tau-lepton candidate multiplied by the absolute value of the pseudorapidity\\
\hline
MET  & transverse missing energy \\
\hline
$\Delta \phi\left(\tau_{\mathrm{h}}, \mathrm{MET}\right)$  & azimuthal angle between the hadronic tau-lepton candidate and the transverse missing energy direction \\
\hline
$M\left(\tau_{\mathrm{h}}, \mathrm{jets}\right)$  & invariant mass of the selected jets and the hadronic tau-lepton\\
\hline
$M_{T}\left(\tau_{\mathrm{h}}, \mathrm{MET}\right)$  & transverse mass of the hadronic tau-lepton candidate and transverse missing energy\\
\end{longtable}
\end{center}
\noindent The aplanarity and sphericity account for the spherical topology of the top-quark decay products and the $q \times \left|\eta\left(\tau_{\mathrm{h}}\right)\right|$ variable exploits the charge-symmetry of $\mathrm{t}\bar{\mathrm{t}}$ events in contrast to W + jets events.
\\
\\
The total number of $\mathrm{t}\bar{\mathrm{t}}$ signal events amounts to 43,570 and the number of QCD background events amounts to 11,176. The data samples are split into a training set of 20,000 $\mathrm{t}\bar{\mathrm{t}}$ signal events and 5000 QCD background events; the remaining events are used for validation. For the training of the neural network classifier the \textsc{PyTorch} package~\cite{NEURIPS2019_9015} is used. A feed-forward neural network architecture with two hidden layers, ReLU activations and a sigmoid function in the last layer has been chosen. The standard binary cross-entropy (\textsc{bce}) loss function is used with a batch size of 256. The input to the training are the eight high-level features described above. The features have been rescaled to have a mean of zero and a standard deviation of one. 
\begin{figure}[h]
\begin{subfigure}[c]{0.49\linewidth}
\centering
\includegraphics[width=1\linewidth]{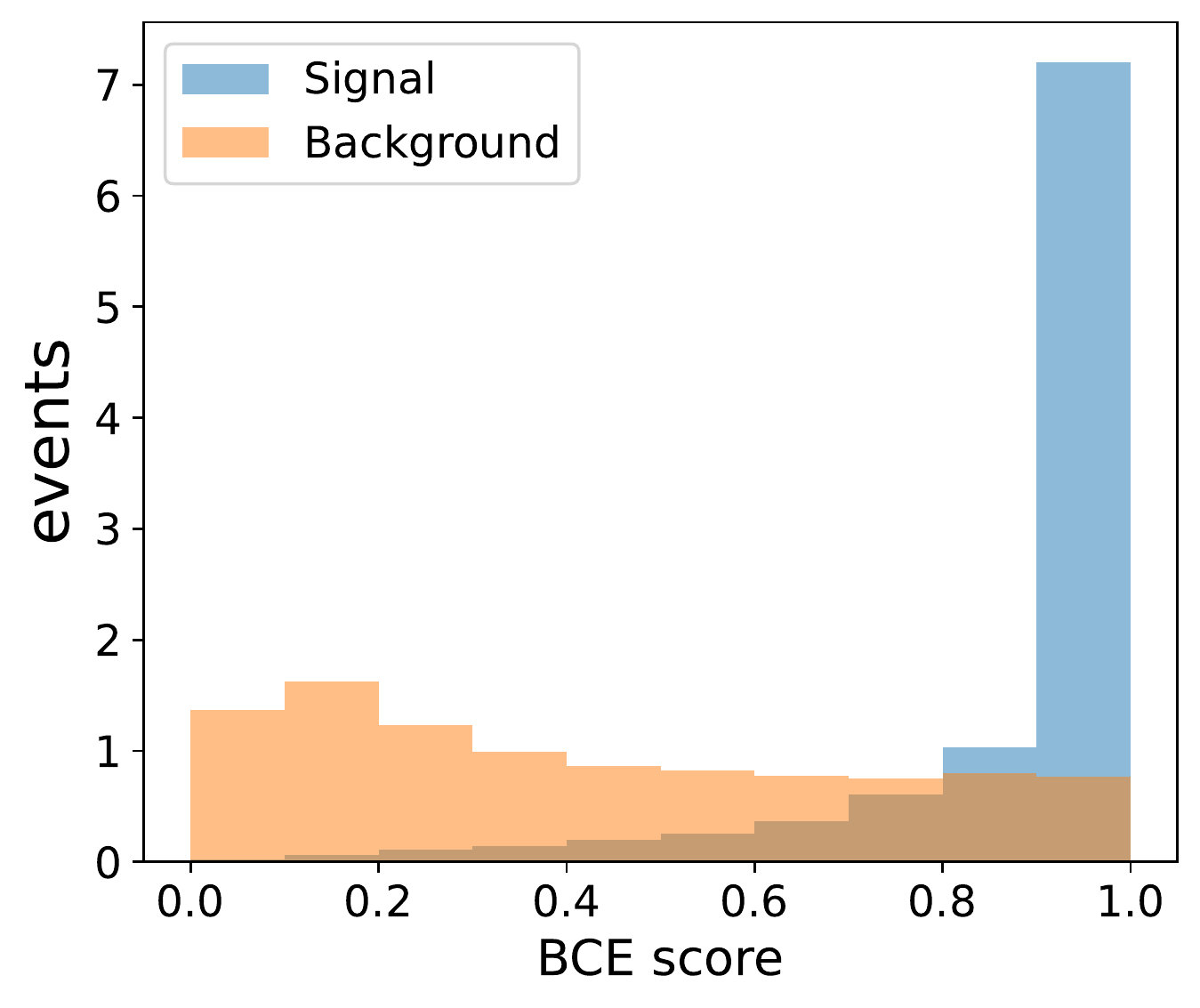}
\end{subfigure}
\hfill
\begin{subfigure}[c]{0.49\linewidth}
\centering
\includegraphics[width=1\linewidth]{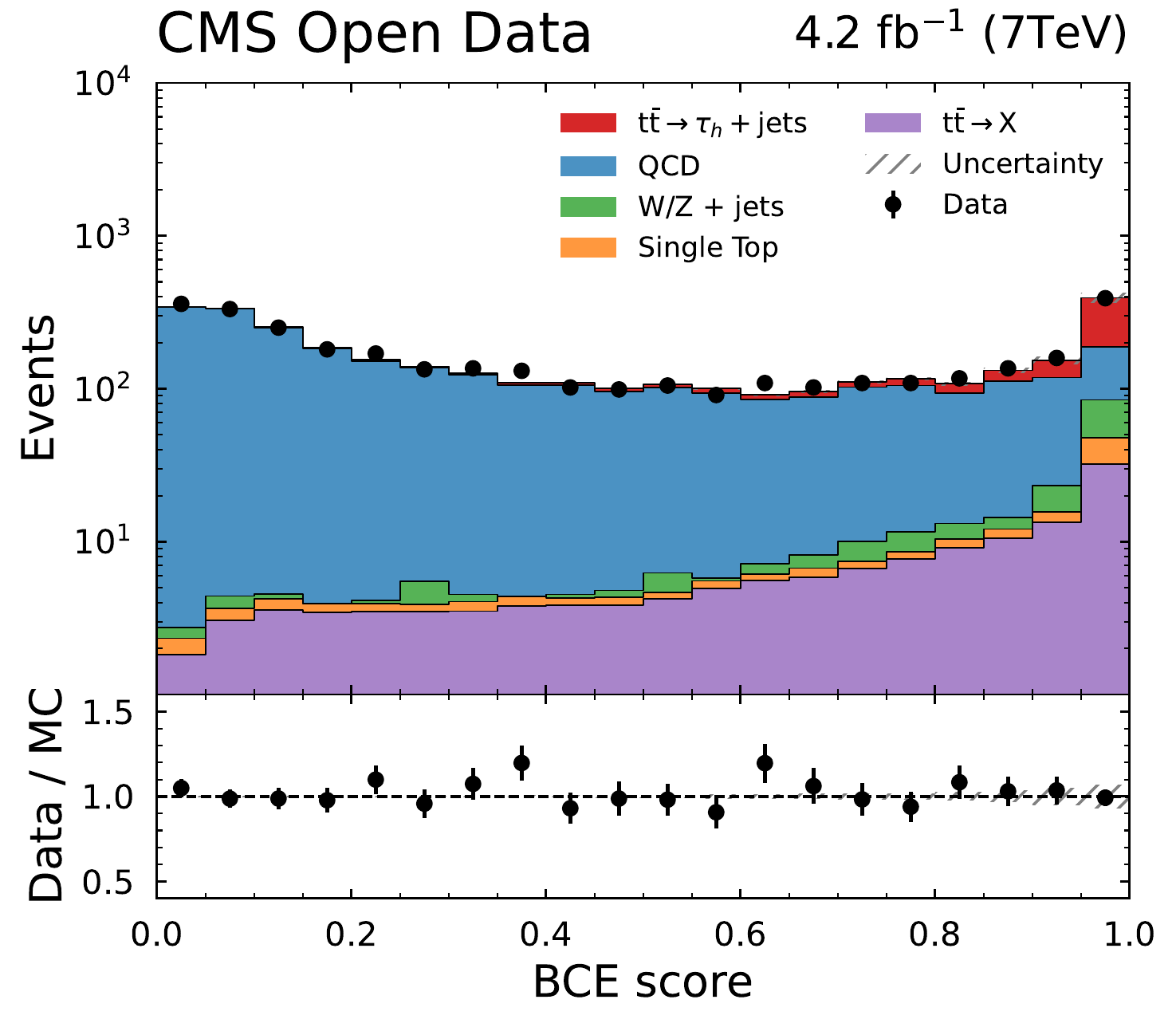}
\end{subfigure}
\caption{Left panel: classifier score of a neural network trained with binary cross-entropy. Right panel: data-simulation agreement for the neural network classifier score.}
\label{bce}       
\end{figure}
For the final \textsc{bce} model, $20$ neurons per layer and a learning rate of $0.001$ were chosen. The model is trained for 100 epochs with the ADAM optimiser; weights that give the lowest \textsc{bce} loss on the validation set are stored.
The binned classifier score of the neural network model is shown in the left panel of Fig.~\ref{bce} for the validation set. Application of the NN to CMS Open Data produces the data-simulation agreement shown in the right panel. The normalisation of the data-driven QCD sample is set to the best fit value after performing a simple log-likelihood fit. A good agreement between data and simulation is observed.

\subsection{Systematic uncertainties}
The main sources of systematic uncertainty are those due to uncertainty in the Jet Energy Scale $(\mathrm{JES})$, the PDF variations, and the tau energy scale. The calculation of systematic uncertainties follows the description of the original CMS analysis, but uses the detailed recipes recommended for the legacy CMS Open Data. 
The uncertainty corresponding to the JES is estimated by shifting the jet energy up and down by the uncertainties corresponding to one standard deviation. For the systematic effect of Jet Energy Resolution uncertainties the distribution of jet energies is smeared by one standard deviation. The corrections are propagated to the missing transverse energy measurement, following the recommended recipes in the CMS Open Data description.
The uncertainty of the tau energy correction is evaluated by shifting the value of the tau energy up and down by \mbox{$\pm$ 3\%}, following the original analysis. The uncertainty of the hadronically-decaying tau identification efficiency is estimated to be $6 \%$~\cite{ttjets}. 
The uncertainty due to the application of the b-tagging scale factors for b-, c- and light-jets to the simulated events is estimated by shifting the value of the applied scale factors by the uncertainty corresponding to one standard deviation. For the b-mistagging reweighting method on the multijet data sample the uncertainty is estimated to 5\%, following the choice of the original analysis. The estimation of the statistical uncertainty corresponding to the trigger efficiency for the particle-flow jets and the particle-flow tau is done by recalculating the trigger weight with a $\pm1 \sigma$ statistical variation of the efficiencies of the jets and the tau for all signal and background processes. 
Following the original analysis, a $\pm$ 5\% uncertainty is accounted for the tau-leg trigger efficiency measurement, in order to take into account the fact that the used reference sample to estimate the tau trigger efficiency consists mainly of jets misidentified as tau-lepton candidates. 
The uncertainty from the association of the matrix elements to the parton showers is estimated to be $3 \%$~\cite{ttjets}. 
The uncertainty of the choice of PDFs on the signal acceptance is estimated by adding in quadrature the $2 \times 22$ reference PDFs associated to CTEQ6.  
The uncertainties in the cross sections for the different simulated background processes are taken from theoretical calculations used in the original analysis.
The uncertainty coming from the top-quark mass is evaluated to a $3 \%$ relative uncertainty in the measured cross section and the dependence on the renormalisation and factorisation scales is estimated for the $\mathrm{t}\bar{\mathrm{t}}$ processes to be $2 \%$~\cite{ttjets}. The uncertainty on the luminosity measurement is estimated to be $2.2 \%$.

\subsection{Cross section measurement}

The original CMS analysis estimates the QCD multijet background and the $\mathrm{t}\bar{\mathrm{t}}$ signal fraction with a negative maximum likelihood fit to the classifier output distribution. Minor backgrounds are subtracted from the data prior to the fit. Systematic uncertainties are calculated by repeating the fit with templates varied by $\pm 1 \sigma$ of the respective systematic source. However, to be consistent with modern LHC analyses, for the CMS Open Data analysis the inference is done with a profile likelihood fit based on the \textsc{cabinetry} package~\cite{refId0} and the CMS \textsc{combine} tool~\cite{combine}, which takes into account all systematic variations simultaneously.
To test a hypothesised value of the signal strength $\mu$ the following profile likelihood ratio is used:
\begin{equation}
t_{\mu} = -2 \ln \frac{L(\mu, \hat{\hat{\boldsymbol{\theta}}}(\mu))}{L(\hat{\mu}, \hat{\boldsymbol{\theta}})}
\end{equation}
where $\hat{\hat{\boldsymbol{\theta}}}(\mu)$ refers to the conditional ML estimators of $\boldsymbol{\theta}$ given a strength parameter $\mu$, which means that it maximises $L$ for a given value of $\mu$. The denominator is the maximised (unconditional) likelihood function, {\em i.e.}, $\hat{\mu}$, and $\hat{\boldsymbol{\theta}}$ are their ML estimators. 
\begin{figure}[h]
\begin{subfigure}[c]{0.6\linewidth}
\centering
\includegraphics[width=1\linewidth]{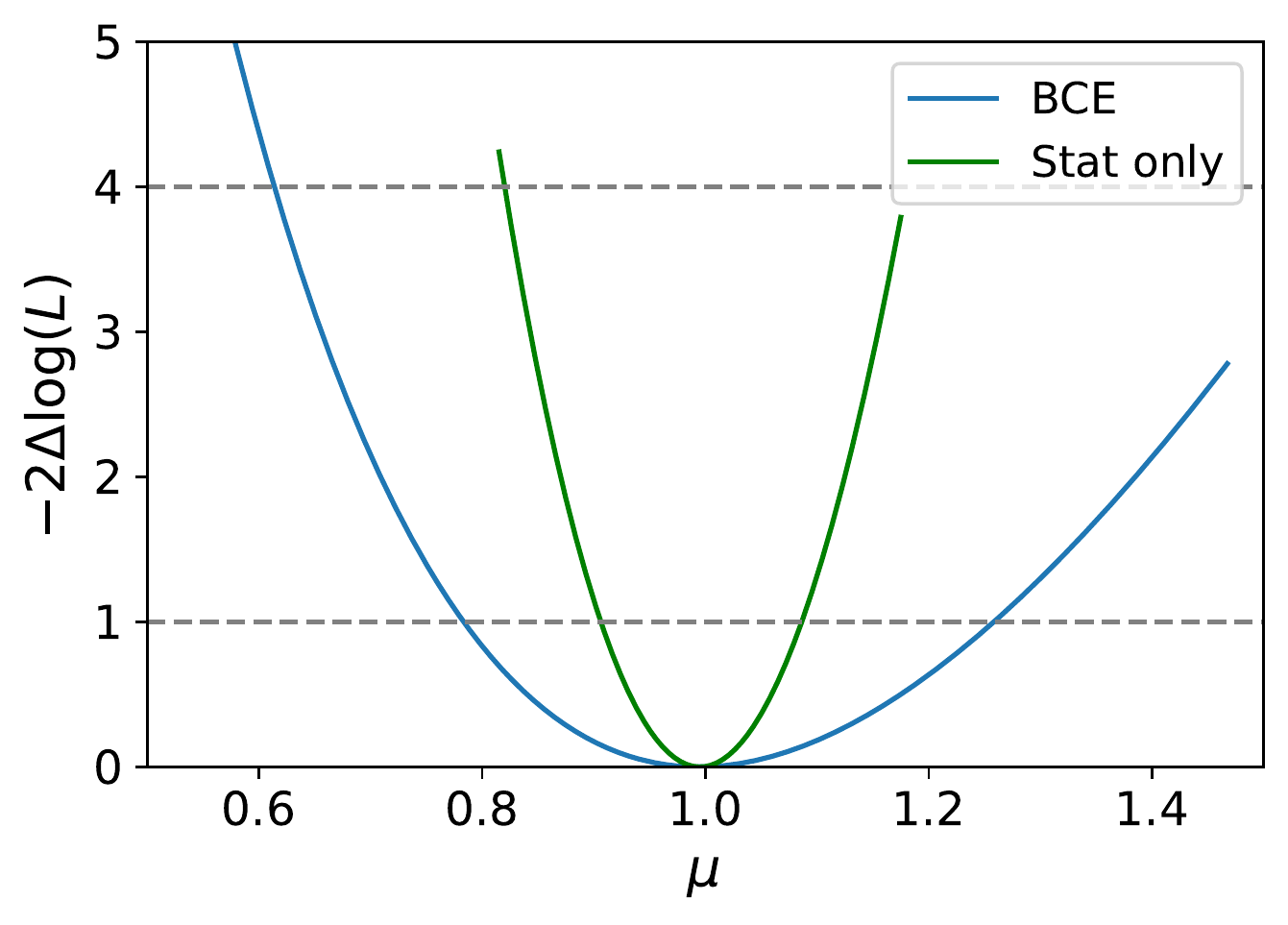}
\end{subfigure}
\hfill
\begin{subfigure}[c]{0.4\linewidth}
\centering
\includegraphics[width=1\linewidth]{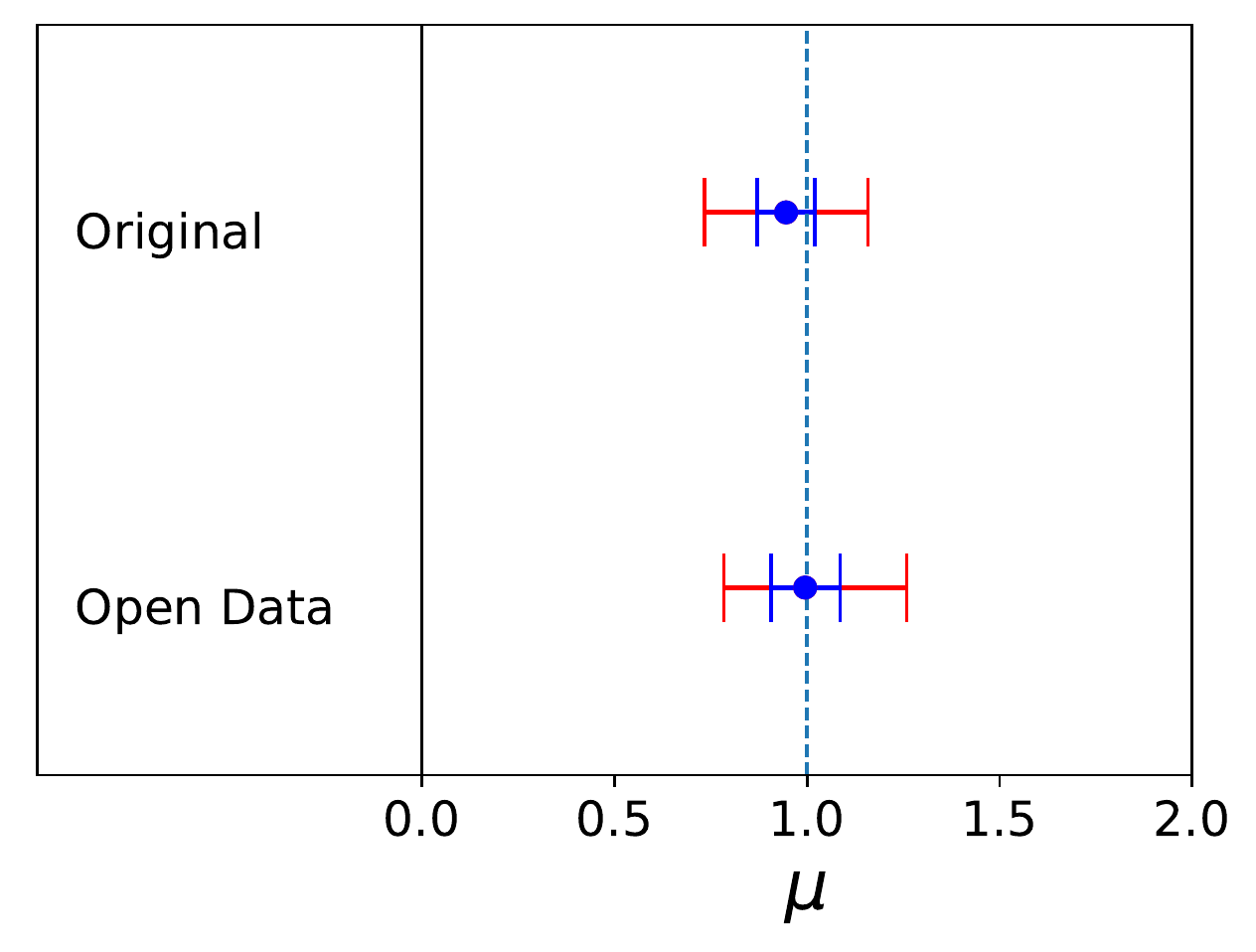}
\end{subfigure}
\caption{Left panel: profile likelihood scan for the signal strength $\mu$ with the summary statistic obtained by training a neural network classifier with binary cross-entropy. Right panel: comparison of the signal strength $\mu$ between the original analysis and the reproduced analysis with CMS Open Data. Blue bars indicate the size of statistical uncertainties.}
\label{ll_scan}       
\end{figure}
Shape variations of the templates are taken into account by using morphing techniques. 
On Asimov data, the signal strength with systematic uncertainties evaluates to:
\begin{equation}
\mu = 1.00^{+0.24}_{-0.19}~\mathrm{(syst.)} \pm 0.08~\mathrm{(stat.)}
\end{equation}
and on the observed data the signal strength evaluates to:
\begin{equation}
\mu = 0.99^{+0.25}_{-0.19}~\mathrm{(syst.)} \pm 0.09~\mathrm{(stat.)}
\end{equation}
An explicit scan of the profile likelihood, is shown in Fig.~\ref{ll_scan}.
The original analysis quotes a value of:
\begin{equation}
\mu = 0.94 \pm0.08 ~\mathrm{(stat.)} \pm 0.19 ~\mathrm{(syst.)}
\end{equation}
The result obtained with the CMS Open Data analysis is in good agreement with the original one. Systematic uncertainties are slightly larger for the CMS Open Data analysis, which can be explained by the different recipes used to calculate the uncertainties. In conclusion, the reproduced analysis with CMS Open Data is a realistic measurement and thus can be used to study the \textsc{inferno} algorithm in a real world example.

\section{Application of INFERNO}\label{sec:3}

In the following the adaptation of the \textsc{inferno} algorithm to a typical HEP problem will be described and its performance will be evaluated based on the measurement of the $\mathrm{t}\bar{\mathrm{t}}$ production cross section in the $\tau$+jets channel in pp collisions at $\sqrt{s}=7$ TeV obtained with CMS Open Data described in the previous section.

\subsection{Extension of INFERNO to HEP Data}

The original code for the \textsc{inferno} algorithm described in Ref.~\cite{DECASTRO2019170} has been developed for the structure of a synthetic problem. However, systematic uncertainties typically encountered in HEP have a special structure and thus the algorithm needs to be adapted to HEP-like systematics. 
Importantly, systematics in HEP are often given as $\pm 1 \sigma$ variations, which result in an overall distortion in the shape of the  variables~\cite{conway2011incorporating}. An example is a Jet Energy Scale uncertainty that shifts all jet energies in an event in the same direction. Distortions of this type can be modelled by changing parameters (like the energy scale) in the MC simulation and recalculating the shifted distributions. For example, raising and lowering the energy scale by one standard deviation, and recalculating the distributions, leads to three measures of the shape and normalisation of the distributions, respectively at minus one sigma, central, and plus one sigma. Another common situation are systematic uncertainties that are given as event weights corresponding to a $+1\sigma$ and $-1\sigma$ variation. In this case the nominal distribution can be weighted event-by-event to obtain three measures of the shape (and normalisation) of the distributions. 
The three measures of the spectral shape can be converted into a continuous estimate in each bin as a function of the energy scale factor by introducing a “morphing” parameter. This has been first implemented in the \textsc{PyTorch} version of \textsc{inferno}~\cite{gilesstrong_2021_5040810}, which also reproduces the results of the synthetic problem. Moreover an alternative version of an interpolation algorithm that is used in the standard fitting tool of CMS, \textsc{combine}~\cite{combine}, has been added, as well as functions that can deal with asymmetric normalisation uncertainties and an alternative differentiable summary statistic. By making use of the interpolation technique, and a suitable preprocessing of the systematic variations, the \textsc{inferno} algorithm has been extended to run with an arbitrary number of HEP-like systematics. The code for this algorithm has been published in~\cite{lukas_layer_2022_6080791}.

\subsection{Training and Inference with one Nuisance Parameter}

\begin{figure}[h]
\begin{subfigure}[c]{0.5\linewidth}
\centering
\includegraphics[width=1\linewidth]{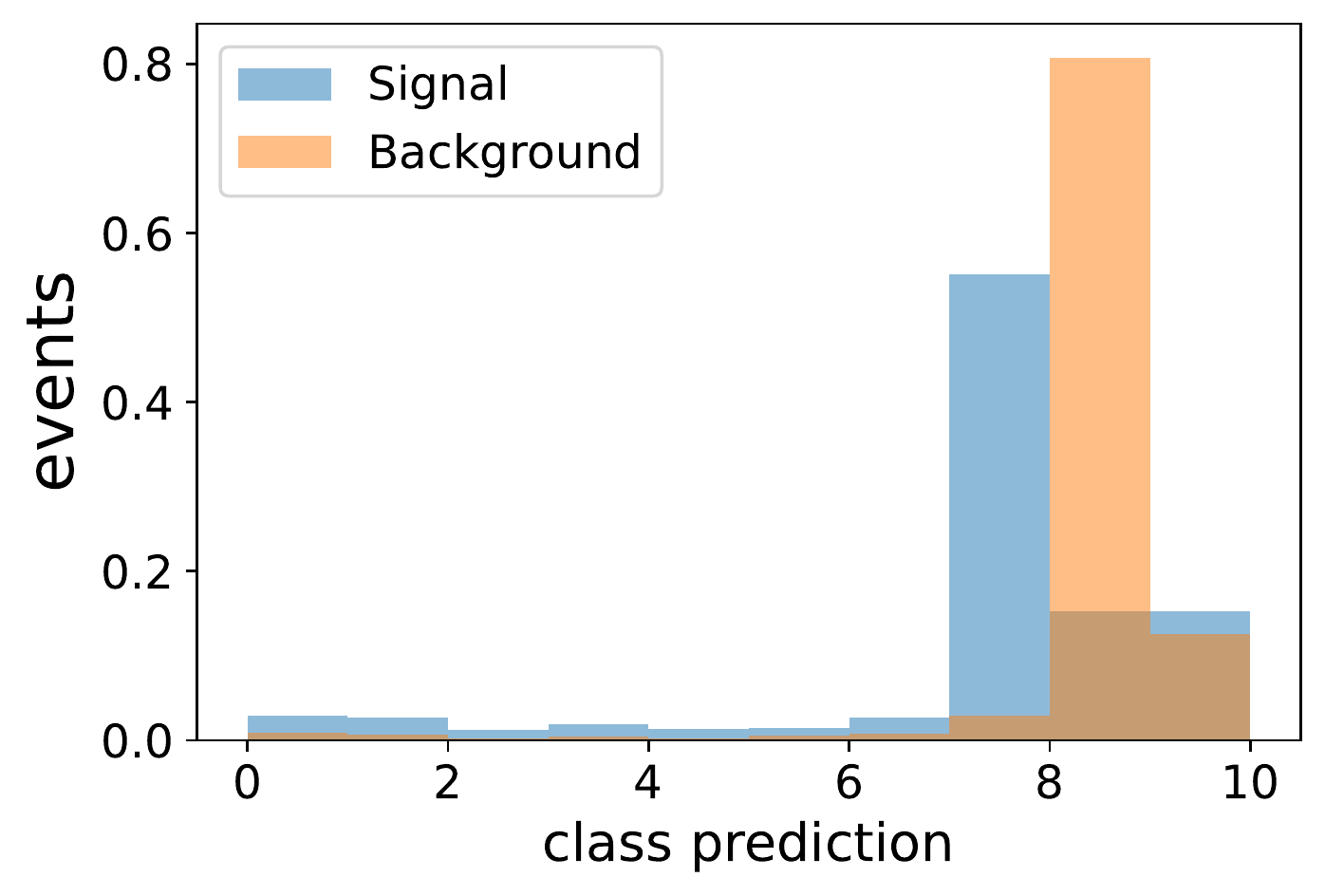}
\end{subfigure}
\hfill
\begin{subfigure}[c]{0.5\linewidth}
\centering
\includegraphics[width=1\linewidth]{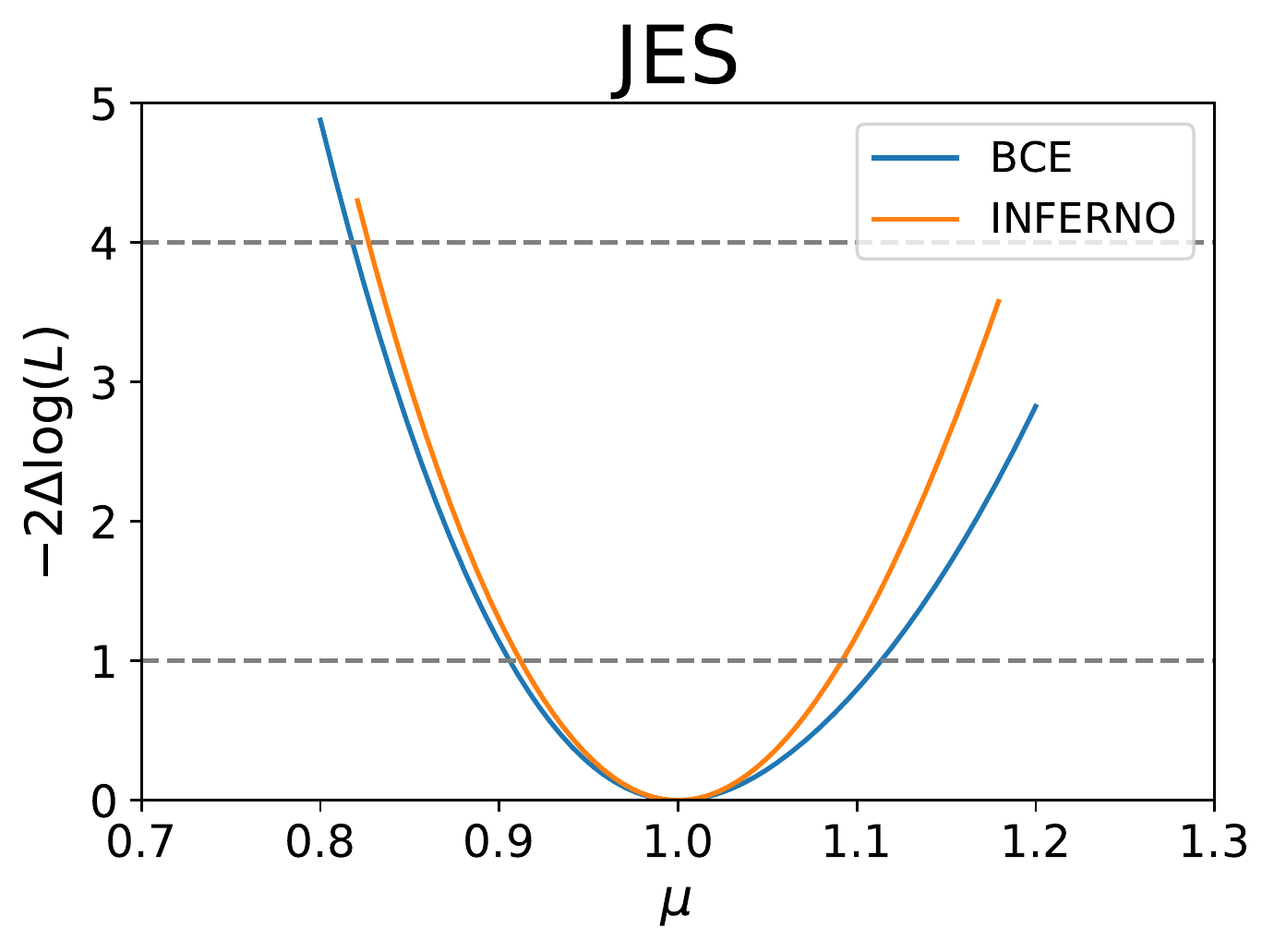}
\end{subfigure}
\caption{Left panel: class prediction of the \textsc{inferno} model trained with the JES variation.
Here the \textit{classes} are not, {\em e.g.} signal and background, but rather the bins of the summary statistic that is learnt by the network: no ordering of the fractional population of signal and background in these bins is to be expected.
Right panel: profile likelihood scan for \textsc{inferno} trained with the JES variation.}
\label{inferno_train}       
\end{figure}

In order to quantify the behaviour of the \textsc{inferno} algorithm applied to the analysis described in Sec.~\ref{sec:2}, several studies based on artificial systematic shifts that affect the shape of the classifier, as well as the normalisation, have been performed. It was concluded that \textsc{inferno} has the potential to mitigate the impact of nuisance parameters that affect the shape of the classifier. This is investigated below by taking the example of a training with one of the most important nuisance parameters, the Jet Energy Scale. An \textsc{inferno} model consisting of a two-layer neural network with $60$ neurons, a learning rate of $0.001$ and a temperature $\tau$ of $0.1$ is trained for 100 epochs. The effect of the Jet Energy Scale variation is included in the \textsc{inferno} training by interpolating between the $\pm 1\sigma$ variations. Thus, the \textsc{inferno} model includes two parameters: the number of expected signal events $s$ and the nuisance parameter corresponding to the Jet Energy Scale $\theta^{\mathrm{JES}}$. The variance of $s$ is chosen as the loss value. The \textsc{inferno} model is compared to an optimised \textsc{bce} model trained on the same dataset. The approximate covariance matrix can be calculated during the training of the \textsc{bce} model by binning the model predictions and using the \textsc{inferno} algorithm to calculate the inverse of the Fisher information matrix.
\begin{figure}[ht]
  \centerline{\includegraphics[width=.8\textwidth]{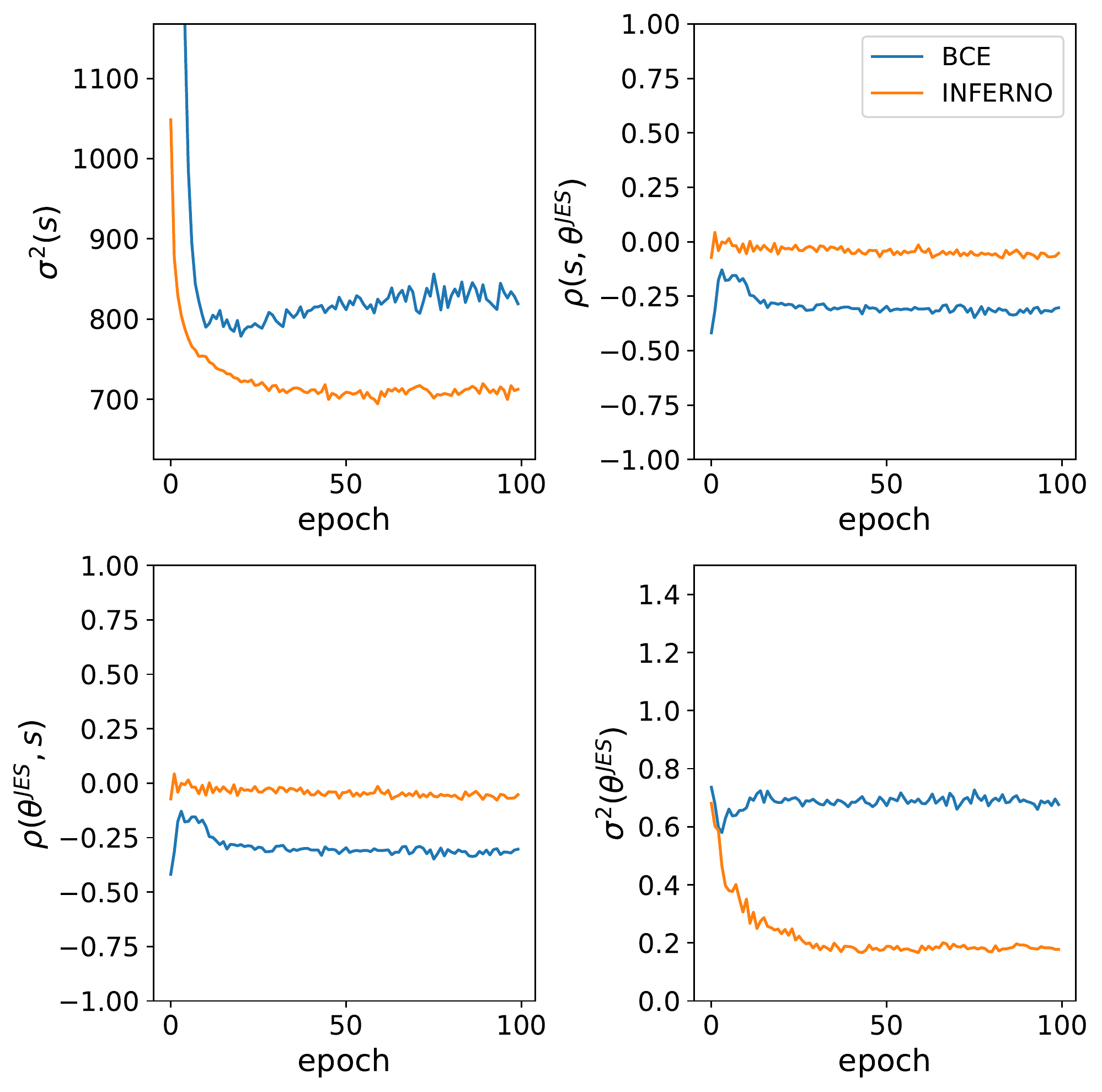}}
  \caption{Evolution of the covariance matrix evaluated on the validation set during training. The diagonal elements (top left and bottom right panels) show the variance of the expected number of signal events $s$ and the variance of the JES nuisance parameter $\theta^{\mathrm{JES}}$. The off-diagonal elements show the correlation coefficient $\rho(s, \theta^{\mathrm{JES}})$.}
  \label{cov_jes}
\end{figure}
Figure~\ref{cov_jes} shows the evolution of the approximated $2\times2$ covariance matrix, monitored after each epoch on the validation set. The diagonal elements of the figure display the variance of the expected number of signal events $\sigma^2(s)$ and the variance of the nuisance parameter $\theta^{\mathrm{JES}}$, denoted by $\sigma^2(\theta^{\mathrm{JES}})$. The off-diagonal elements show the correlation coefficient $\rho$ between $s$ and $\theta$, which is defined as the covariance of the variables divided by the product of their standard deviations:
\begin{equation}
\rho(\mu, \theta^{\mathrm{JES}})=\frac{\operatorname{Cov}(\mu, \theta^{\mathrm{JES}})}{\sigma(\mu) \sigma(\theta^{\mathrm{JES}})} ~.
\end{equation}
Evaluating the evolution of the covariance matrix shows that the variance of the parameter of interest $s$ converges to a lower value with the \textsc{inferno} model compared to the classifier trained with \textsc{bce}.
The correlation coefficient $\rho$ between $s$ and $\theta^{\mathrm{JES}}$ converges to a value closer to zero with the \textsc{inferno} model. 
The variance of $\theta^{\mathrm{JES}}$ converges to a lower value with \textsc{inferno} compared to the \textsc{bce} classifier. This indicates that the \textsc{inferno} algorithm makes optimal use of the data in order to decorrelate the parameter of interest $s$ from the nuisance parameter $\theta^{\mathrm{JES}}$, which results in a lower variance for $s$ compared to a model trained with \textsc{bce}.
The class predictions for the validation set for the output of the \textsc{inferno} training are shown in the left panel of Fig.~\ref{inferno_train}.
\\
\\
The summary statistics produced in the training with \textsc{inferno} are then used in a profile likelihood fit implemented in the \textsc{cabinetry}~\cite{refId0} package. Only the nuisance parameter $\theta$ corresponding to the Jet Energy Scale variation is included in the fit, such that the same conditions as in the \textsc{inferno} training are given. 
The \textsc{minos} algorithm~\cite{minos} is used to calculate the 68\% confidence interval, while the uncertainty of the nuisance parameter and the correlation coefficient is calculated from the \textsc{hesse} estimate. The measured confidence interval for the signal strength $\mu$ evaluated on Asimov data based on the \textsc{inferno} and \textsc{bce} summary statistics is:
\begin{equation}
\begin{array}{ll}
\mu_{\mathrm{\textsc{bce}}}^A = 1.000^{+0.113}_{-0.094}   \\  \\
\mu_{\mathrm{\textsc{inf}}}^A = 1.000^{+0.091}_{-0.087}
\end{array},
\end{equation}
the post-fit value of the nuisance parameter $\theta$ obtained from the \textsc{hesse} estimate is:
\begin{equation}
\begin{array}{ll}
\theta^A_{\mathrm{\textsc{bce}}} = 0.000 \pm 0.905  \\
\theta^A_{\mathrm{\textsc{inf}}} = 0.000 \pm 0.602 
\end{array}
\end{equation},
and the correlation coefficient $\rho$ between $\mu$ and $\theta$ has been evaluated to:
\begin{equation}
\begin{array}{ll}
\rho^A_{\mathrm{\textsc{bce}}} = -0.42  \\
\rho^A_{\mathrm{\textsc{inf}}} = -0.10 ~.
\end{array}
\end{equation}
A scan of the profile likelihood for the parameter $\mu$ is shown in the right panel of Fig.~\ref{inferno_train}, both for the \textsc{inferno} and \textsc{bce} model. 

The above results illustrate that the \textsc{inferno} summary statistic yields a narrower confidence interval for the signal strength $\mu$ compared to the \textsc{bce} model if a \textit{ShapeNorm} nuisance parameter is present. It further shows that the estimates of the covariance matrix in the \textsc{inferno} training are good estimates of the values obtained from the fit of the Asimov data. As  observed during the training, the \textsc{inferno} algorithm reduces the correlations between $\mu$ and $\theta$. This suggests that the main improvement obtained with \textsc{inferno} is due to reduced correlations between the POI and the nuisance parameters.

\subsection{Training and Inference with all Nuisance Parameters}

\begin{figure}[ht]
  \centerline{\includegraphics[width=.8\textwidth]{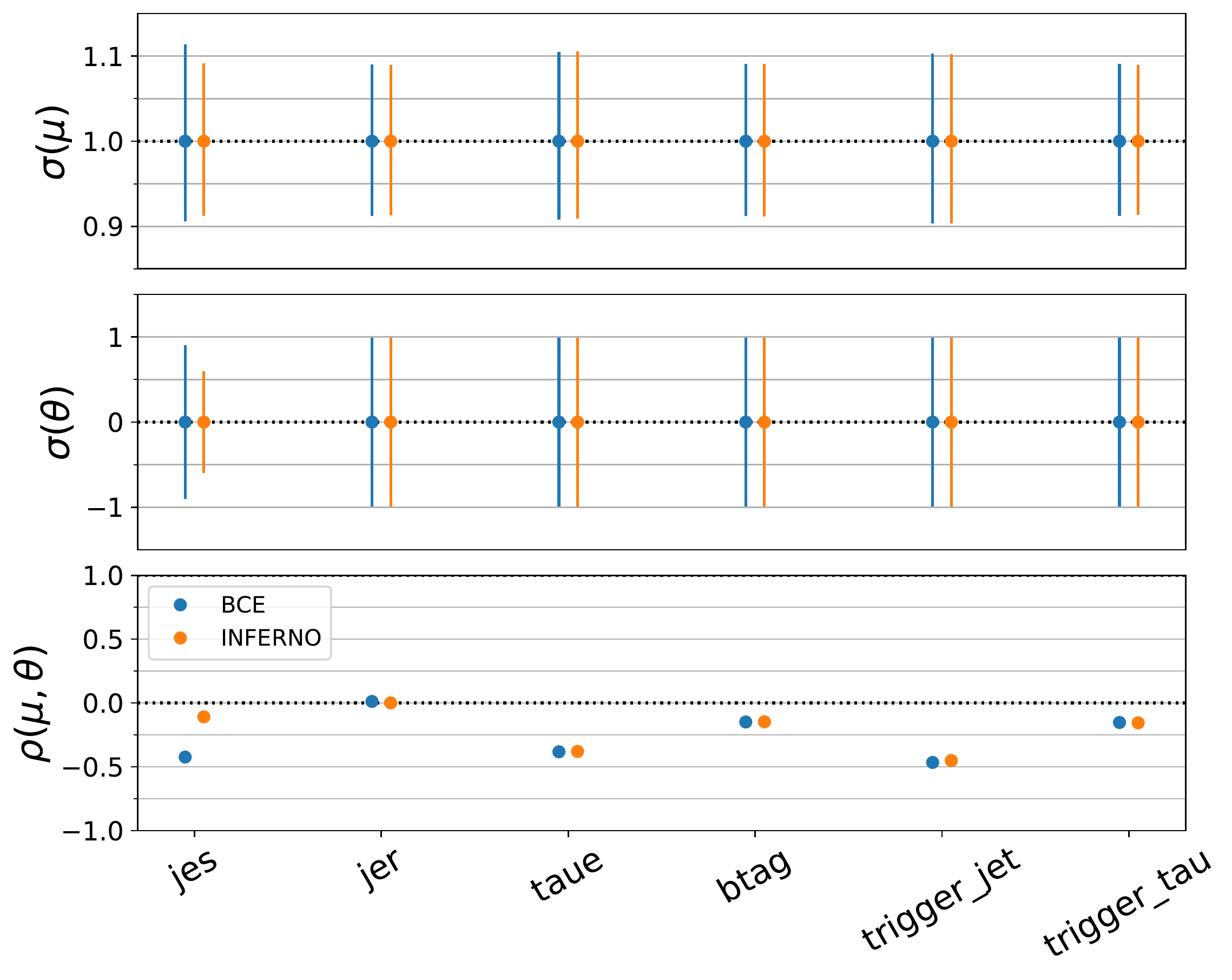}}
  \caption{Comparison of the confidence intervals for $\mu$ (top panel), uncertainty of the respective \textit{ShapeNorm} nuisance parameter $\theta$ (middle panel) and the correlation coefficient $\rho$ obtained in a profile likelihood fit for each of the considered \textit{ShapeNorm} nuisance parameters.}
  \label{comp_sn}
\end{figure}

The study for the Jet Energy Scale variation described in the previous section has been repeated for all relevant systematic uncertainties that affect the shape and normalization of the classifier. In Fig.~\ref{comp_sn} the \textsc{minos} uncertainty for $\mu$, the uncertainty of the corresponding nuisance parameter $\theta$ and the correlation coefficient $\rho$ is shown.
Comparing the results between the \textsc{inferno} and \textsc{bce} model shows that an improvement is mainly possible for the JES nuisance parameter, where \textsc{inferno} manages to decorrelate it from the signal strength $\mu$. For the other nuisance parameters the obtained confidence intervals and correlations are very similar. These nuisance parameters have only a small influence on the shape of the classifier. Thus \textsc{inferno} cannot decorrelate these parameters from the POI, and hardly any improvement with \textsc{inferno} over \textsc{bce} is obtained. \\
\\
For the complete cross section measurement, a model is trained that takes all relevant \textit{ShapeNorm} nuisance parameters into account. A hyperparameter scan verified that an \textsc{inferno} model consisting of a two-layer neural network with $60$ neurons, a learning rate of $0.001$ and a temperature $\tau$ of $0.1$ trained for 100 epochs is an appropriate choice.
\begin{figure}[h]
\begin{subfigure}[c]{0.5\linewidth}
\centering
\includegraphics[width=1\linewidth]{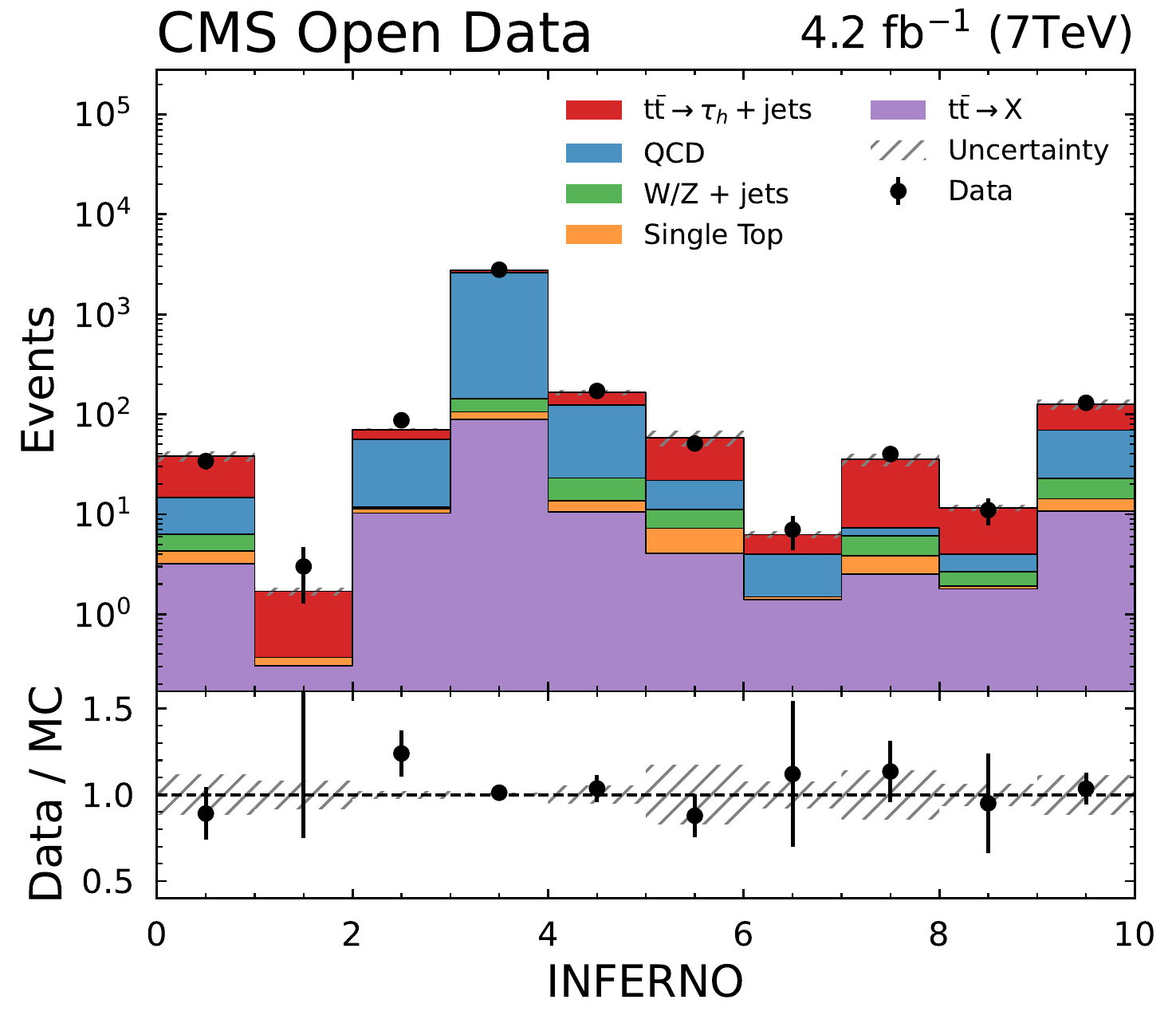}
\end{subfigure}
\hfill
\begin{subfigure}[c]{0.5\linewidth}
\centering
\includegraphics[width=1\linewidth]{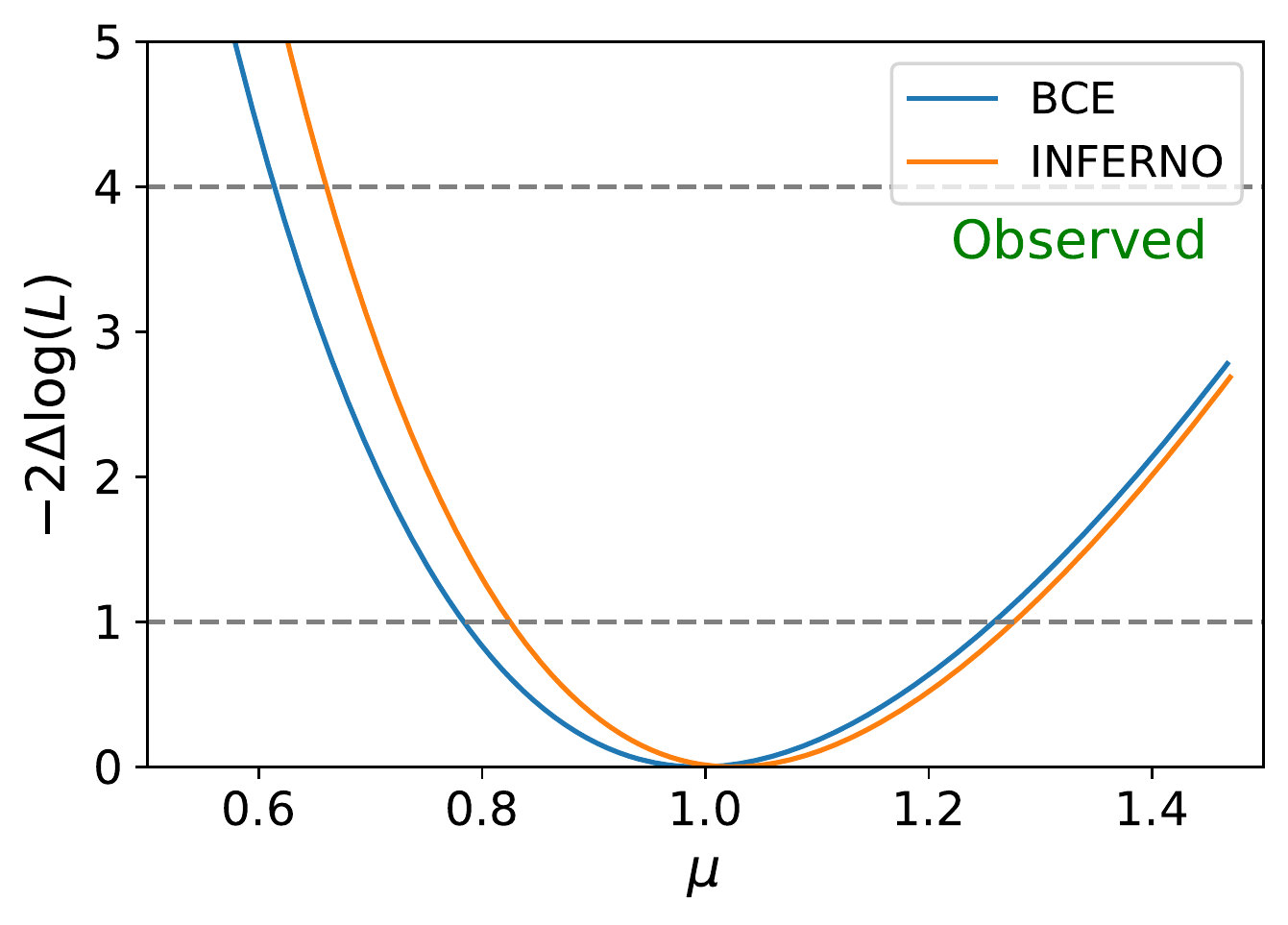}
\end{subfigure}
\caption{Left panel: data-simulation agreement for the \textsc{inferno} model trained with all relevant \textit{ShapeNorm} nuisance parameters. Right panel: comparison of the profile likelihood scan for a fit to the observed data for the \textsc{inferno} and \textsc{bce} model with all relevant nuisance parameters included.}
\label{final}       
\end{figure}
The data-simulation agreement for the \textsc{inferno} output is shown in the left panel of Fig.~\ref{final}. A good agreement between data and simulation is observed. In order to compare with the result obtained for the signal strength $\mu$ in Sec.~\ref{sec:2}, the full profile likelihood fit that includes all relevant nuisance parameters for all relevant processes is performed with \textsc{cabinetry}. This makes the assumption that the effect of the minor backgrounds were negligible for the training of \textsc{inferno} and the obtained summary statistic is still optimal. \\
The resulting confidence interval for $\mu$ based on the \textsc{inferno} summary statistic evaluated on Asimov data is measured to:
\begin{equation}
\mu^A_{\textsc{inf}} = 1.00^{+0.22}_{-0.17}~\mathrm{(syst.)} \pm 0.09~\mathrm{(stat.)}
\end{equation}
and the resulting confidence interval for $\mu$ evaluated on the observed data is measured to:
\begin{equation}
\mu_{\textsc{inf}} = 1.02^{+0.23}_{-0.18}~\mathrm{(syst.)} \pm 0.09~\mathrm{(stat.)} ~.
\end{equation}
The profile likelihood scan is shown in the right panel of Fig.~\ref{final}. The likelihood scan obtained with the classifier trained with \textsc{bce} in Sec.~\ref{sec:2} is included for comparison.
The central values measured for the signal strength $\mu$ are in good agreement. A moderate improvement in the precision of the confidence interval is obtained with the \textsc{inferno} summary statistic compared to the \textsc{bce} summary statistic. The improvement is due to the mitigation of the Jet Energy Scale nuisance parameter. This result shows that \textsc{inferno} has the potential to improve the precision of confidence intervals if nuisance parameters affect the shape of the classifier by reducing the correlations between the POI and the relevant nuisance parameters.

\section{Conclusions}

In this work a published CMS analysis has been fully reproduced with CMS Open Data, which can serve as a benchmark for similar studies in the future. This allowed us to adapt the \textsc{inferno} algorithm to a realistic HEP problem. Based on the reproduced analysis, a study has been performed that compares inference with summary statistics obtained with \textsc{inferno} to summary statistics obtained by training a model with binary cross-entropy. It has been found that \textsc{inferno} has the potential to mitigate the effect of nuisance parameters that affect the shape of the classifier by decorrelating the signal strength from the nuisance parameters. For the studied CMS Open Data analysis the impact of the JES nuisance parameter has been reduced; only a moderate improvement has been obtained since most of the uncertainties in the analysis affect only the data normalisation. The produced results demonstrate that novel HEP analysis that are affected by systematics that distort the shapes of the distributions can strongly profit from training a summary statistic with \textsc{inferno}. A potential next step in the development of \textsc{inferno} is to apply it in a novel CMS physics analysis. The algorithm can also be extended to take multiple background processes and channels into account. The code for the study described in this paper has been released to the public~\cite{lukas_layer_2022_6080791}.

\clearpage
\section*{Acknowledgements}

The authors thank the CMS experiment and the CERN Open Data Portal Team for making the datasets used in this analysis publicly available. 
This work is supported by the European Innovative Training Network INSIGHTS, funded by the European Union’s Horizon 2020 research and innovation program, call H2020-MSCA-ITN-2017, under Grant Agreement n. 765710. CloudVeneto is acknowledged for the use of computing and storage facilities.

\begin{figure}
\begin{center}
\includegraphics{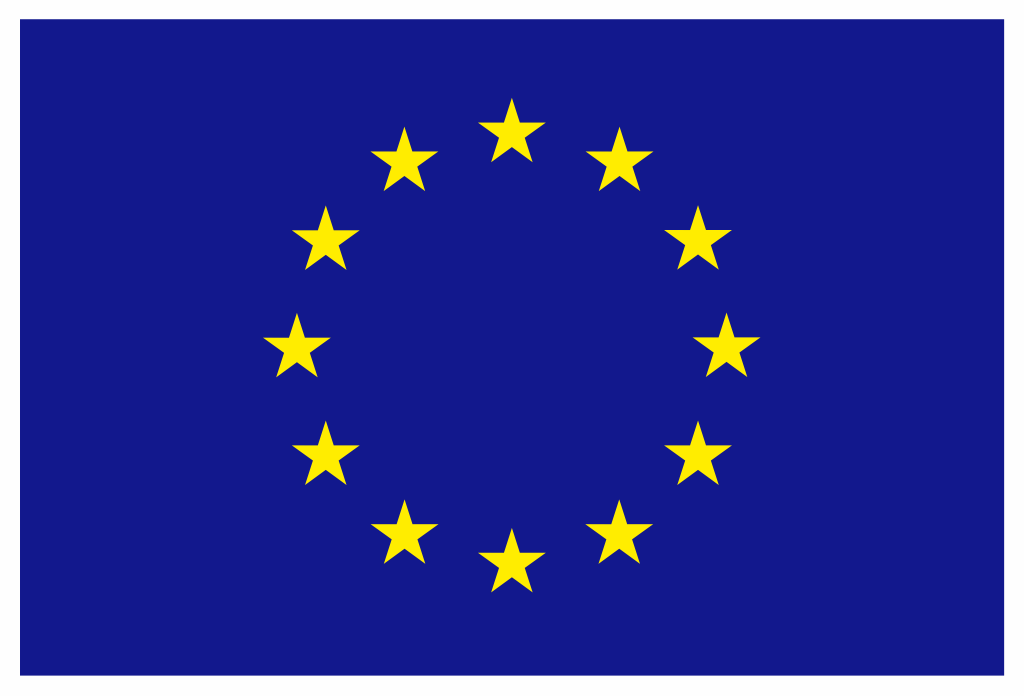}
\end{center}
\end{figure}

\bibliographystyle{unsrt}
\bibliography{referencias}

\end{document}